\documentclass[conference]{IEEEtran}
\usepackage{graphicx}

\begin{document}
\title{Charge Transfer Properties Through Graphene Layers in Gas Detectors}
\author{\IEEEauthorblockN{
P.~Thuiner$^{1,2,*}$,
R.~Hall-Wilton$^{3}$,
R.~B.~Jackman$^{4}$,
H.~M\"uller$^{1}$,
T.~T.~Nguyen$^{4}$,
E.~Oliveri$^{1}$,
D.~Pfeiffer$^{3}$,\\
F.~Resnati$^{1}$,
L.~Ropelewski$^{1}$,
J.~A.~Smith$^{1,4}$,
M.~van~Stenis$^{1}$, and
R.~Veenhof$^{5}$
}
\IEEEauthorblockA{
$^{1}$
CERN, European Organization for Nuclear Research, 1211 Geneva, Switzerland}
\IEEEauthorblockA{
$^{2}$
Technische Universit\"at Wien, Atominstitut, Stadionallee 2, 1020 Vienna, Austria}
\IEEEauthorblockA{
$^{3}$
ESS, European Spallation Source, 22100 Lund, Sweden}
\IEEEauthorblockA{
$^{4}$
London Centre for Nanotechnology and University College London, WC1H 0AH, United Kingdom}
\IEEEauthorblockA{
$^{5}$
Uluda\u{g} University, Department of Physics, 16059 Bursa, Turkey and RD51 Collaboration}
\IEEEauthorblockA{
$^{*}$
e-mail: patrik.thuiner@cern.ch}
}
\maketitle

\begin{abstract}
Graphene is a single layer of carbon atoms arranged in a honeycomb lattice with remarkable mechanical, electrical and optical properties.
For the first time graphene layers suspended on copper meshes were installed into a gas detector equipped with a gaseous electron multiplier.
Measurements of low energy electron and ion transfer through graphene were conducted.
In this paper we describe the sample preparation for suspended graphene layers, the testing procedures and we discuss the preliminary results followed by a prospect of further applications.
\end{abstract}

Graphene, Gas detectors, Gaseous Electron Multiplier, electron transparency, ion transparency

\IEEEpeerreviewmaketitle

\section{Introduction}

Graphene is a semimetal~\cite{Neto:2009}, which conserves~\cite{Zhou:2013} its conducting properties when large areas are supported on a metal. This allows the non-intrusive integration of the material on gold or copper, which constitutes existing detector layers. 

Graphene is the thinnest material to date, consisting of a single layer of carbon atoms arranged in aromatic rings, forming a honeycomb-like structure. One of the most interesting properties is that, despite its atomic thinness, it is very mechanically resistant in the plane of the lattice due to its aromatic rings with strong delocalised double bonds. This allows for stable integration of suspended graphene over the relatively large holes of tens of micrometers in diameter.

These same bonds also allow, to a lesser extent, high mechanical stability in the direction orthogonal to the lattice of the material. The two-dimensional yield strength was measured~\cite{Lee:2008} in 2008 to be 42~Nm$^{-1}$ for a defect-free layer. Similarly the Young’s modulus was found to be about $1.0 \pm 0.1$~TPa as opposed to 200~GPa as for alloyed steels ~\cite{Pavlina:2008} for example. This very high mechanical resistance allows for impermeability to small ions and atoms, all the while allowing electron transparency over certain ranges of energy. 

The basic hexagonal ring has a bond length of 0.142~nm and therefore an inner radius~\cite{Berry:2013} of 0.246~nm. Although this 'gap' is relatively large, the $\pi$ bonds orthogonal to the lattice can be seen as a delocalized cloud of electrons, which overlaps the hole in the hexagon. This reduces the opening pore to be significantly smaller, yielding an effective diameter of 0.064~nm~\cite{Berry:2013}. This is much smaller than the van der Waals radius of most atoms, some of the smallest ones for instance being helium with 0.28~nm and hydrogen with 0.314~nm.
The $\pi$ bonds therefore should allow impermeability to atoms, molecules and ions~\cite{Schedin:2007} assuming they do not have enough energy to go through the electron cloud. Experimentally, suspended graphene has been measured to withstand an irradiation dose up to approximately 10$^{16}$~ions/cm$^2$ at tens of keV energies~\cite{Morin:2012}. Similar experiments have shown that graphene is completely impermeable to Helium atoms up to 6~atm~\cite{Bunch:2008}.

On the other hand, graphene has been shown to exhibit high transparency to electrons with energies ranging from tens of keV up to 300~keV~\cite{Meyer:2007} through a layer of suspended graphene. Although this electron permeability has been shown at high electron energies, its interaction with incoming low energy electrons has not yet been studied.

Preliminary measurements in this direction are done using a gas detector equipped with a Gas Electron Multiplier (GEM)~\cite{Sauli:1997, Bachmann:1999}.

The experimental setup and methods are described in section~\ref{sec:Setup} and section~\ref{sec:Methods}. Results for measurements with single layer graphene are are shown in section~\ref{sec:Single}, those for triple layer graphene in section~\ref{sec:Triple}. Conclusions are presented in section~\ref{sec:conclusions}. Necessary steps for further improvement are expressed in section~\ref{sec:outlooks}, together with comments on possible applications of this technology.

\section{Experimental apparatus}
\label{sec:Setup}
\subsection{The detector}

\begin{figure}[!t]
\centering
\includegraphics[trim={0 0.2cm 0 4.1cm},clip,width=3.4in]{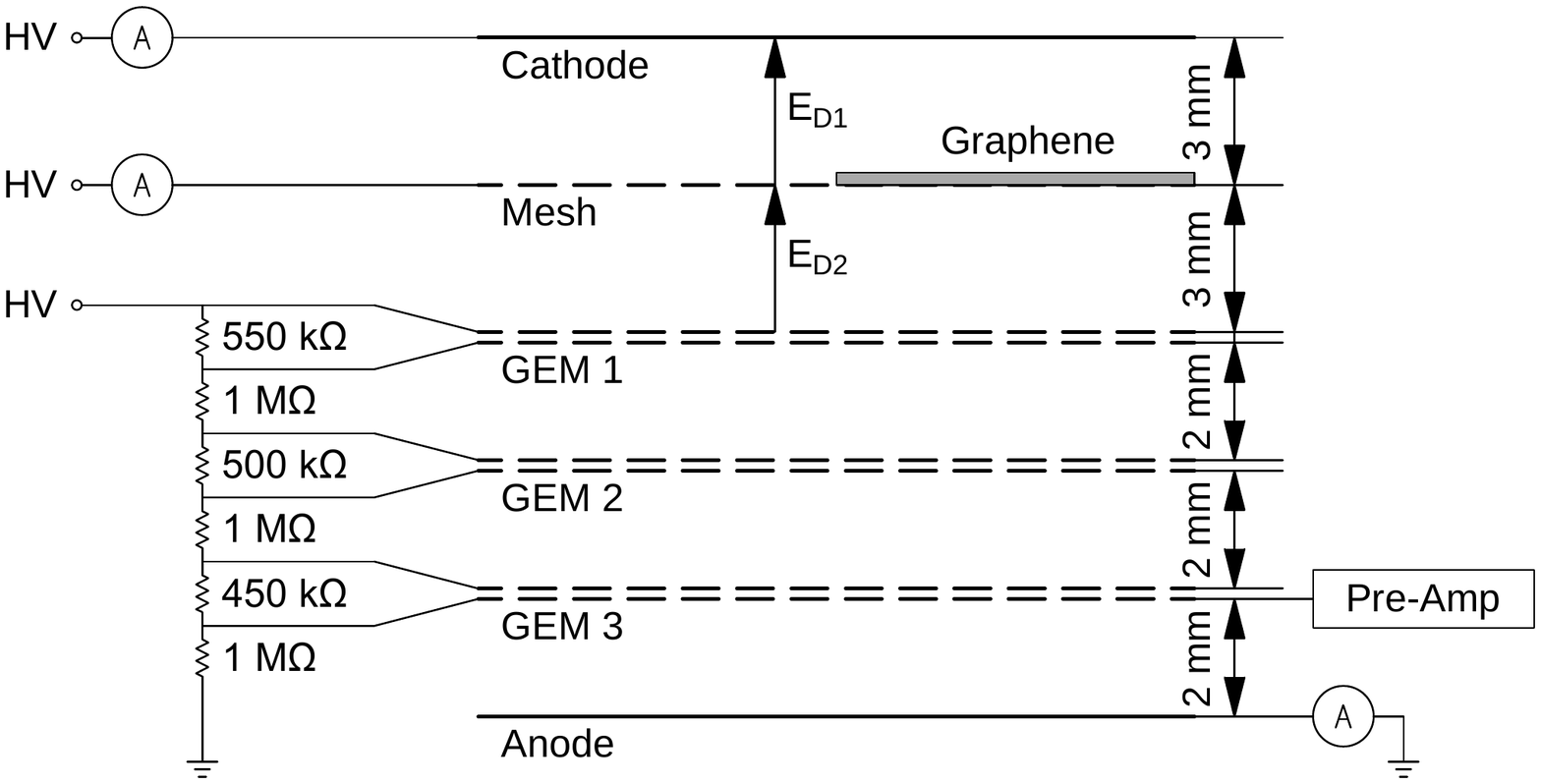}
\caption{Schematic representation of the setup.}
\label{fig:scheme}
\end{figure}

A schematic representation of the detector is shown in Fig.~\ref{fig:scheme}.
It consists of two 3~mm-long conversion volumes separated by a copper mesh covered with the graphene layer.
Two different meshes are used for the studies presented, each 5~$\mu$m thick and $3\times3$~cm$^2$ of area with 30~$\mu$m diameter holes arranged in a honeycomb pattern with a pitch of 60~$\mu$m or 120~$\mu$m respectively.
The graphene occupies only a portion of the mesh, so that comparative measurements with and without the graphene can be performed without modifying the setup.
The conversion volumes are defined on top by the cathode and on the bottom by three $10\times10$~cm$^2$ GEM foils stacked one on top of the other.
The GEMs are powered through a resistor divider. Cathode and mesh are powered individually using a CAEN NDT1471H power supply with 50 pA current resolution.\\
The charges exiting the GEMs are collected onto the anode connected to a Keithley 6487 pico-ammeter.
This configuration allows the simultaneous measurement of the currents at anode, mesh and cathode.
The lowest electrode of the third GEM is connected to a Ortec 142 PC preamplifier and Ortec timing filter amp 474 amplifier to facilitate event by event measurement of the GEM signals. Spectra are taken with an Amp-Tek MCA8000D Pocket MCA.\\
The detector is continuously flushed with an Ar/CO$_2$ gas mixture, using a mass ratio of 90/10 at 5~L/h for the mesh with pitch 120~$\mu$m and a mass ratio of 70/30 at 9~L/h for the mesh with pitch 60~$\mu$m.
Collimated X-rays of 8~keV (approximately 1~mm$^2$ beam size) generated by a copper X-ray gun are used to produce the primary ionisation charges in the conversion volumes. Ionisation electrons from the topmost region must cross the graphene mesh before undergoing multiplication in the GEMs. Similarly, ions produced during the avalanche process in the GEMs must pass the mesh to reach the cathode.
The electron (ion) transparency is defined as the fraction of the charge from the top (bottom) volume reaching the bottom (top).

\subsection{Graphene transfer}
\label{sec:transfer}

Graphene is prepared in monolayer and trilayer forms. The three layers allow the preparation of defect-free layers on copper meshes. 

Graphene is grown as a monolayer on polycrystalline copper by CVD chemistry at 1000$^{\circ}$C with CH$_4$ as a carbon precursor, and H$_2$ and Ar mixtures. The graphene grown on the back of the copper is then removed by immersion in nitric acid.
A layer of 300~nm of PMMA is then spin coated onto the surface of graphene. The copper support is then removed using an aqueous solution of Fe(NO$_3$)$_3$.
The graphene attached to the remaining layer of PMMA is then transferred onto copper meshes.
The PMMA is then removed using a critical point dryer to avoid damaging the single layer of freestanding graphene. 
Trilayer graphene is made using three monolayers transferred on top of each other then layered onto meshes.

\begin{figure}[!t]
\centering
\includegraphics[trim={0 0 0 0},clip,width=3.4in]{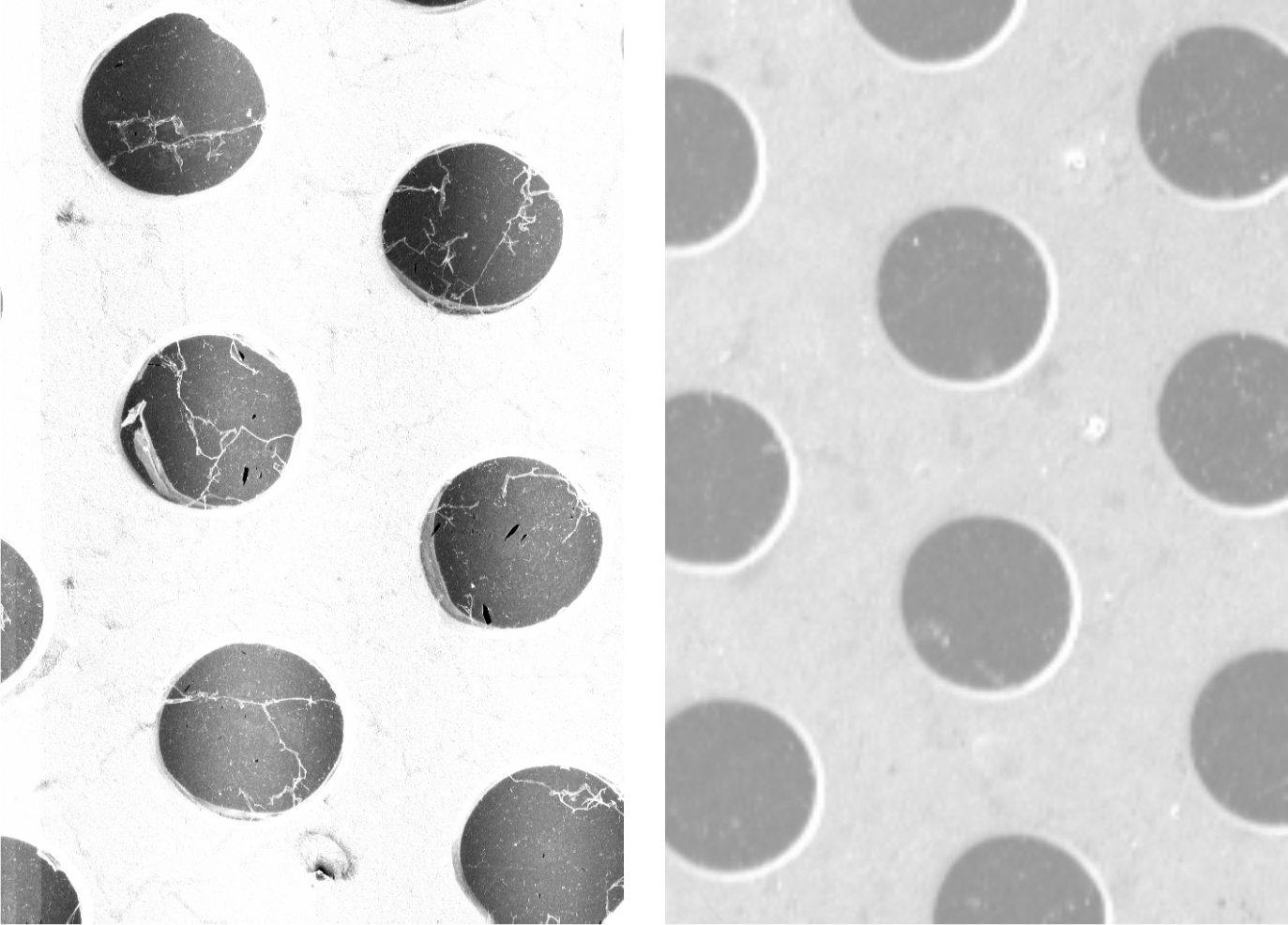}
\caption{SEM images of single layer graphene (left) and triple layer graphene (right) suspended on copper meshes with 30~$\mu$m diameter holes and 60~$\mu$m pitch. Black spots mark gaps, white lines folds in the graphene layer.}
\label{fig:SEM}
\end{figure}

The coverage of the graphene layers on the meshes is assessed using scanning electron microscopy (SEM). The coverage with single layers is found to be $90 \pm 5$\%. Fig.~\ref{fig:SEM} shows the SEM image of a single layer compared to a triple layer with $99.1 \pm 0.5$\% of the holes surface covered. The percentages is determined using an image segmentation technique.

\section{Methods}
\label{sec:Methods}

\begin{figure}[!t]
\centering
\includegraphics[trim={0 0 0 0},clip,width=3in]{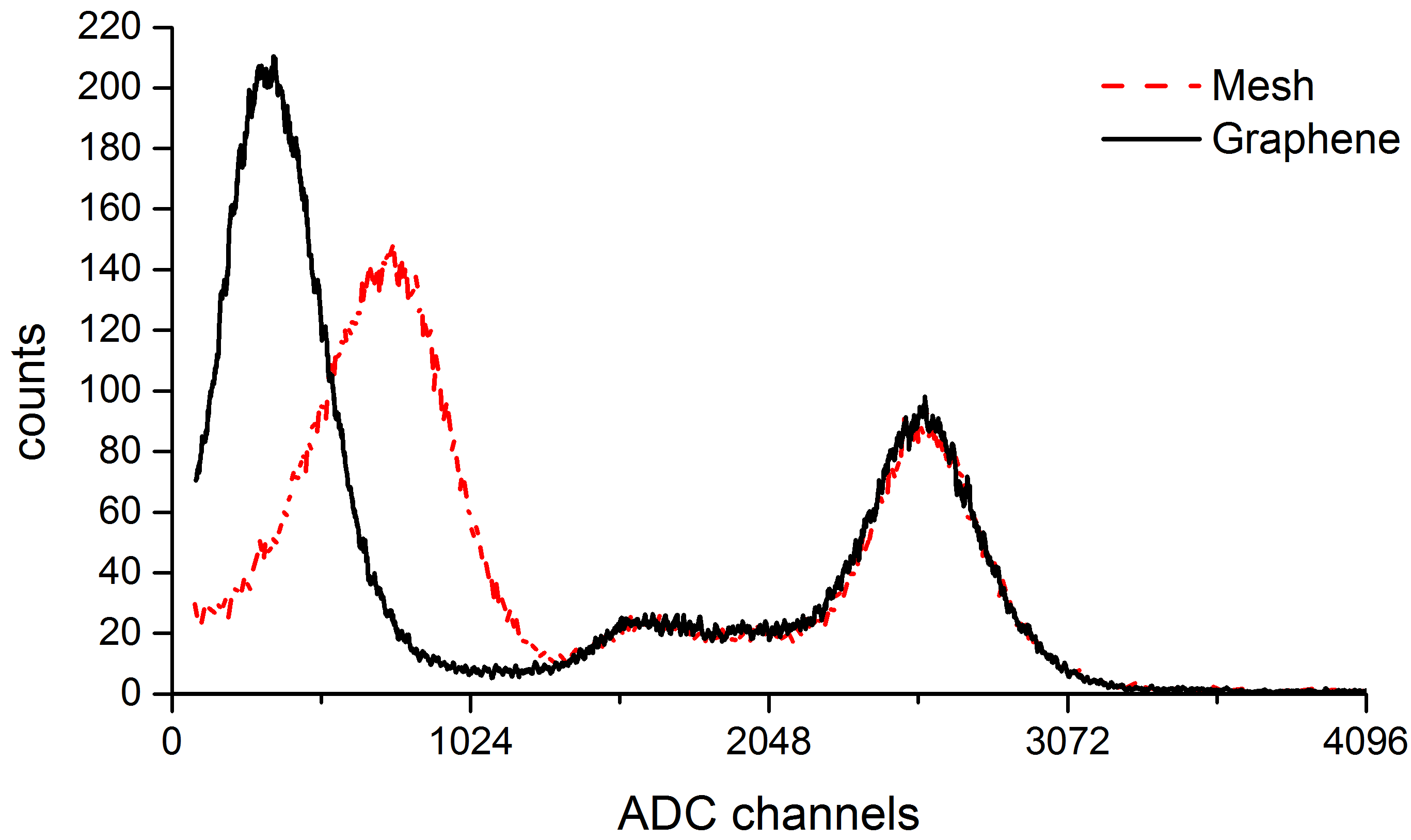}
\caption{Typical pulse height spectra of the GEM signals with (solid) and without (dashed) single layer graphene coverage for $E_{D1} = 20$~V/cm and $E_{D2} = 800$~V/cm in gas mixture of Ar/CO$_2$~70/30}
\label{fig:electronTransSpectra}
\end{figure}

The electron transparency is measured from the pulse height spectrum on exposing the detector to x-rays with an interaction rate in the order of kHz.
Typical spectra are shown in Fig.~\ref{fig:electronTransSpectra}.
While all primary electrons produced below the mesh contribute to the signal of the GEMs, only a fraction produced above the mesh reaches the amplification stage.
Therefore, the peak at around 2500~ADC is due to events from the bottom conversion region and is independent of the presence of the graphene layer on the mesh.
The peak below 1200~ADC is due to events from the top conversion volume.
The ratio of these two peak positions provides an estimation of the electron transparency.

The ion transparency is estimated from the ion currents collected on the cathode ($I_C$) and on the mesh ($I_M$).
Due to the small currents involved, the x-ray interaction rate is increased to the order of $10^5$~Hz.
$I_C+I_M$ normalised to the anode current is checked to be constant for both the covered and uncovered sides.
The ion transparency is then defined as $T_I = I_C/(I_C+I_M)$.

The ratio of the electric fields in the bottom ($E_{D2}$) and in the top ($E_{D1}$) regions governs the transparency of the copper mesh without graphene, i.e. increasing this ratio electrons are more focused into the holes, and vice versa the ions are more focused by decreasing $E_{D2}/E_{D1}$.\\
Further, the electric field configuration has impact on the energy of the drifting electrons: the energy of the electrons in front of the graphene is increased with increasing field $E_{D1}$. Since one expects that the electron transfer probability through graphene increases with the electron energy, increasing $E_{D1}$ should lead to increased electron transparency.

\section{Single Layer Graphene}
\label{sec:Single}

\begin{figure}[!t]
\centering
\includegraphics[trim={0 0 0 0},clip,width=3in]{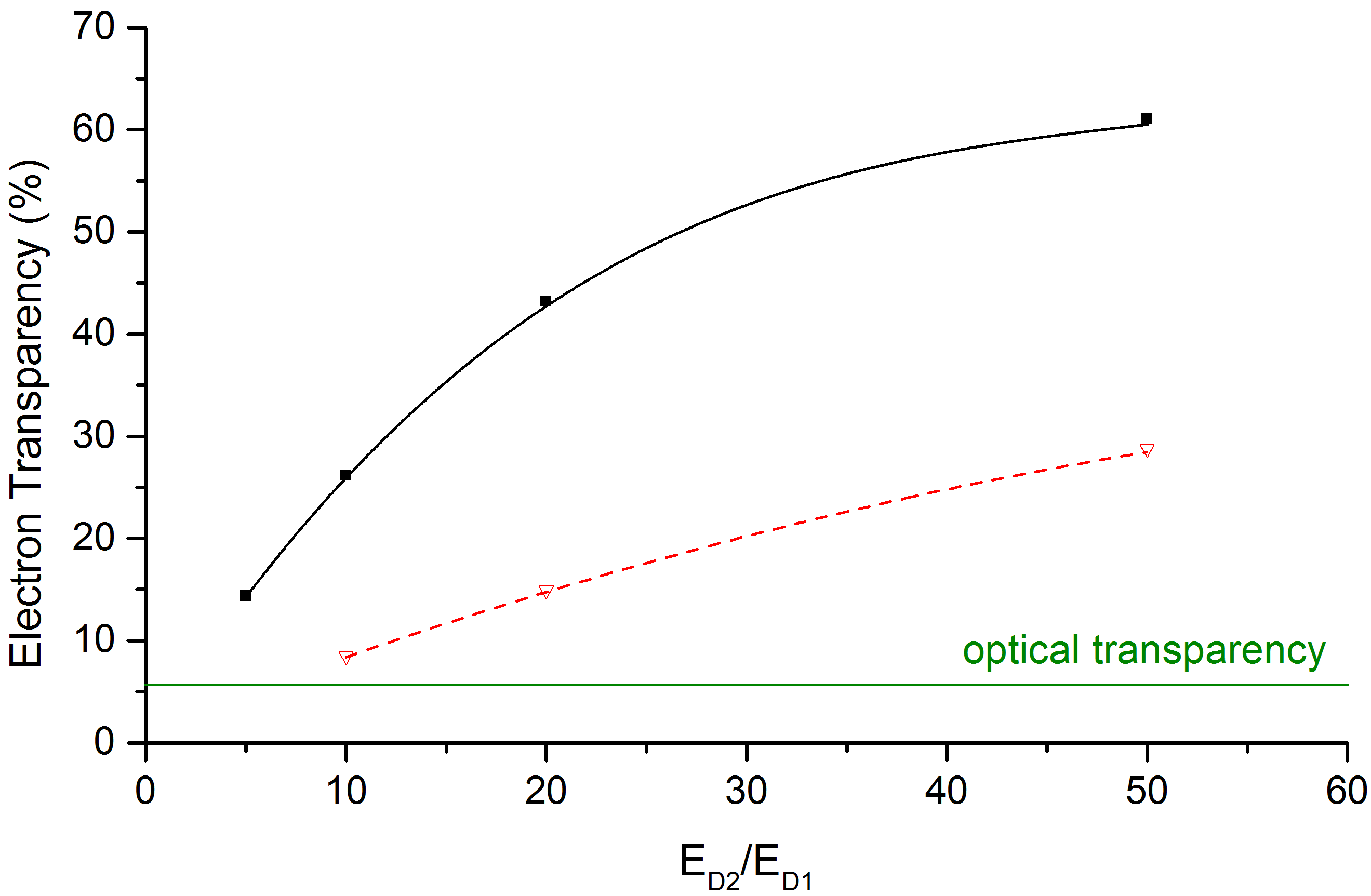}
\caption{Electron transparency for different field ratios $E_{D2}/E_{D1}$ for a mesh with 30~$\mu$m diameter holes and 120~$\mu$m pitch in Ar/CO$_2$~90/10. Optical transparency and guide lines shown for reference.}
\label{fig:electronTransSingle}
\end{figure}

\begin{figure}[!t]
\centering
\includegraphics[trim={0 0 0 0},clip,width=3in]{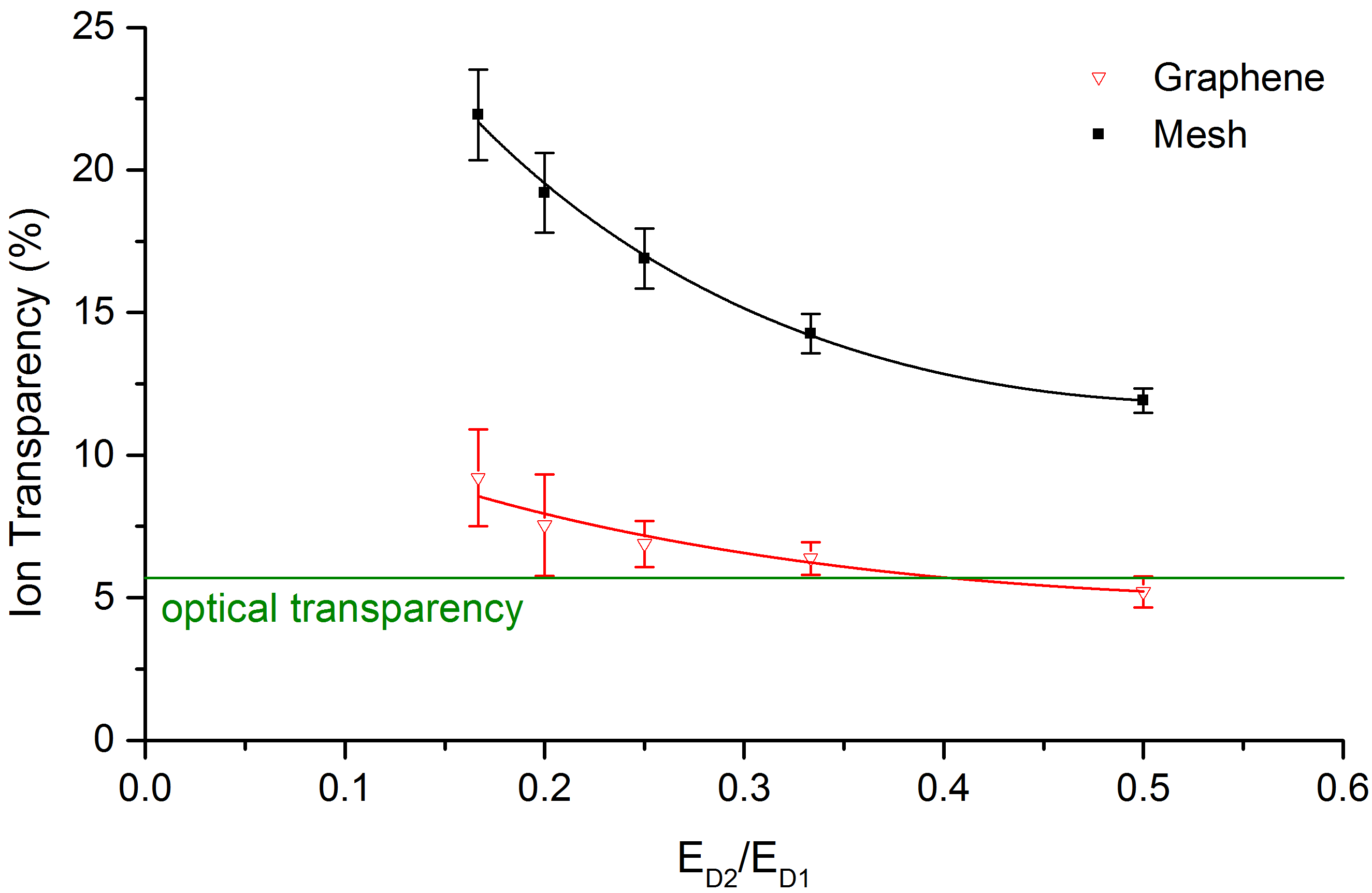}
\caption{Ion transparency for different field ratios $E_{D2}/E_{D1}$ for a mesh with 30~$\mu$m diameter holes and 120~$\mu$m pitch in Ar/CO$_2$~90/10. Optical transparency and guide lines shown for reference.}
\label{fig:ionTransSingle}
\end{figure}

As shown in Fig.~\ref{fig:electronTransSingle} and Fig.~\ref{fig:ionTransSingle} for every tested field configuration studied the transparency of graphene both for electrons and ions is decreased by approximately a factor of two compared to the bare mesh. The transparency of the graphene shows a dependence on the field configuration, analogous to the normal behaviour of the uncovered mesh.\\
Assuming an undamaged graphene layer, with a continuous conducting flat surface, field line focussing is not expected to occur and this effect should only appear on an uncovered mesh. 
Under such conditions, the highest fraction of electrons and ions passing the mesh is expected to be equal or lower than the optical transparency of the mesh. However, as plotted on Fig.~\ref{fig:electronTransSingle} and Fig.~\ref{fig:ionTransSingle} the measured electron and ion transparency exceeds the amount allowed by the optical transparency at field ratios of $E_{D2}/E_{D1} \geq 10$ or $E_{D2}/E_{D1} \leq 0.5$ respectively. While electron transparencies lower than 2\% were observed for lower field ratios, they were not resolvable from the noise.

The partial permeability of ions, the electron and ion transparencies exceeding the optical transparency of the mesh, and the dependence on field ratios showing a comparable trend to an uncovered mesh strongly suggest charge transfer through defects in the layer. Therefore, further measurements with triple layer graphene are conducted to disentangle the contribution from damages and the effective transparency of the graphene layer itself.

\section{Triple Layer Graphene}
\label{sec:Triple}

\begin{figure}[!t]
\centering
\includegraphics[trim={0 0 0 0},clip,width=3in]{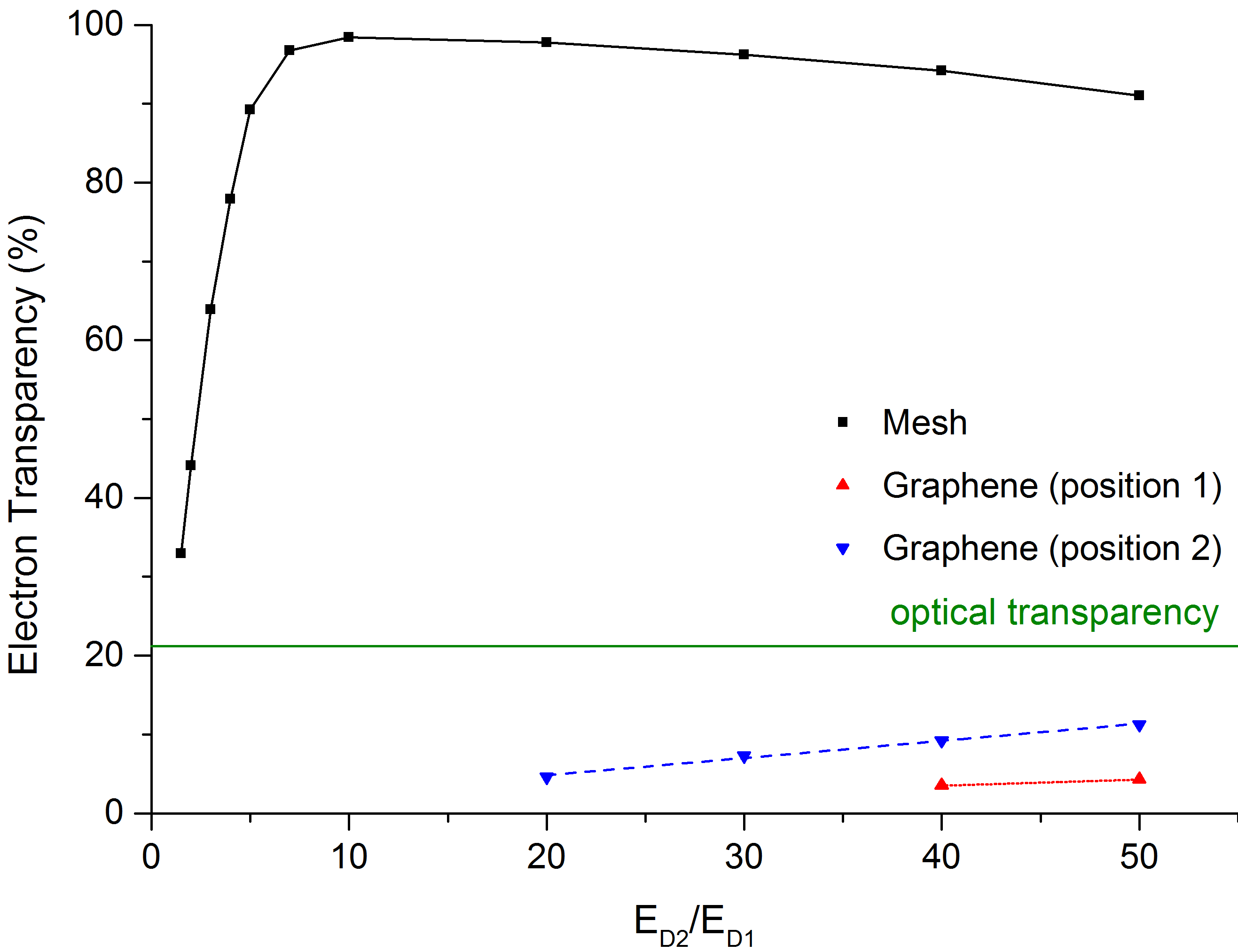}
\caption{Electron transparency of triple layer graphene for different $E_{D2}/E_{D1}$ with $E_{D2}=2000$~V/cm. Graphene suspended on a mesh with 30~$\mu$m diameter holes and 60~$\mu$m pitch. Detector operated in Ar/CO$_2$~70/30. Optical transparency and guide lines shown for reference.}
\label{fig:electronTranspTriple}
\end{figure}

\begin{figure}[!t]
\centering
\includegraphics[trim={0 0 0 0},clip,width=3in]{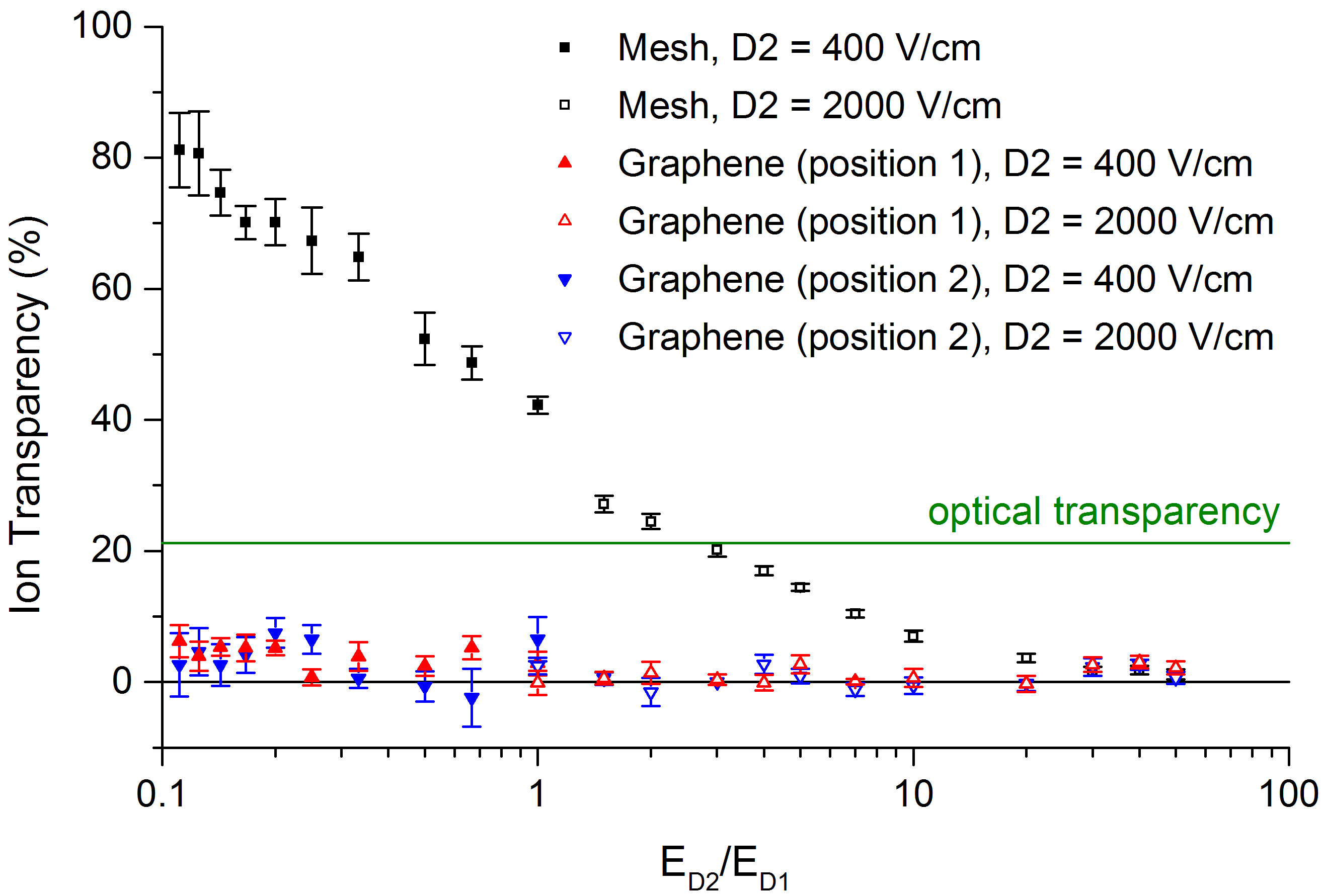}
\caption{Ion transparency of triple layer graphene for different $E_{D2}/E_{D1}$ with $E_{D2}=400$~V/cm. Graphene suspended on a mesh with 30~$\mu$m diameter holes and 60~$\mu$m pitch. Detector operated in Ar/CO$_2$~70/30. Optical transparency shown for reference.}
\label{fig:ionTransTriple}
\end{figure}

Triple layer graphene is transferred with the methods described in section \ref{sec:transfer}. 
The x-ray beam size is reduced to approximately 0.2~mm$^2$ to increase sensitivity to layer inhomogeneities and defects, while at the same time maintaining a flux comparable to single layer graphene measurements.

Two points on the layer (approximately 1.4~mm apart) are investigated, as shown in Fig.~\ref{fig:electronTranspTriple}. The electron transparency of the uncovered mesh reaches more than 90\% for $E_{D2}/E_{D1}>5$ with a loss of transparency at higher ratios due to loss of primary electrons at very low fields $E_{D1}$.
The difference in maximum transparency of the bare mesh between the triple layer and single layer tests is attributed to the difference in optical transparency of the meshes used.
\\
For the points investigated on the graphene, the electron transparency increases with increasing ratio $E_{D2}/E_{D1}$. 
The sensitivity of the measurements didn't allow to measure transparencies lower than 2\%.
At large $E_{D2}/E_{D1}$, transparencies reached 10\% for one position and 5\% for another.

Fig.~\ref{fig:ionTransTriple} shows the ion transparency of the uncovered mesh reaching approximately 80\% for very low $E_{D2}/E_{D1}$. Both positions on the graphene layer yield around or less than 5\% ion transparency, with errors of the same order of magnitude. A clear trend towards lower $E_{D2}/E_{D1}$ is not observable.

Despite the fact that disuniformities of the electron transparency on the graphene layer may be due to defects or varying layer thicknesses as a results of defects not propagating through all three layers the asymmetry in the changes of electron and ion transparency 
moving from the uncovered mesh to graphene suggests that the measurement technique is reaching the required sensitivity to measure the intrinsic charge transfer properties of the graphene itself.

\section{Conclusions}
\label{sec:conclusions}

For the first time graphene layers were used in combination with a gas detector to measure its permeability to electrons and ions at low energies. The measurements performed prove the techniques developed are suitable even though the evaluation of the graphene transparency is complicated by sample defects.
The measurements require high quality layers with uniform overall coverage and layer thickness and very low number of defects. Studies with single layer graphene showed behaviour similiar to a mesh, suggesting charge transfer mainly through defects in the graphene.\\
Graphene triple layers were produced and effectively transferred. Very high overall coverage of the holes was achieved. While there were still contributions from defects, the results hint to an asymmetry between electron and ion transfer.

\section{Outlooks}
\label{sec:outlooks}
In the future, the role of the graphene defects will be addressed by correlating micrometer-scale SEM mapping with charge transfer measurements.
In addition, further improvements on the layer quality will be achieved by refining the transfer procedures. Studies for production of graphene covered meshes and GEMs without the need for a transferring process are ongoing. If successful, these methods may be applicable to larger scales.\\

\begin{figure}[!t]
\centering
\includegraphics[trim={0 0 0 0},clip,width=2.8 in]{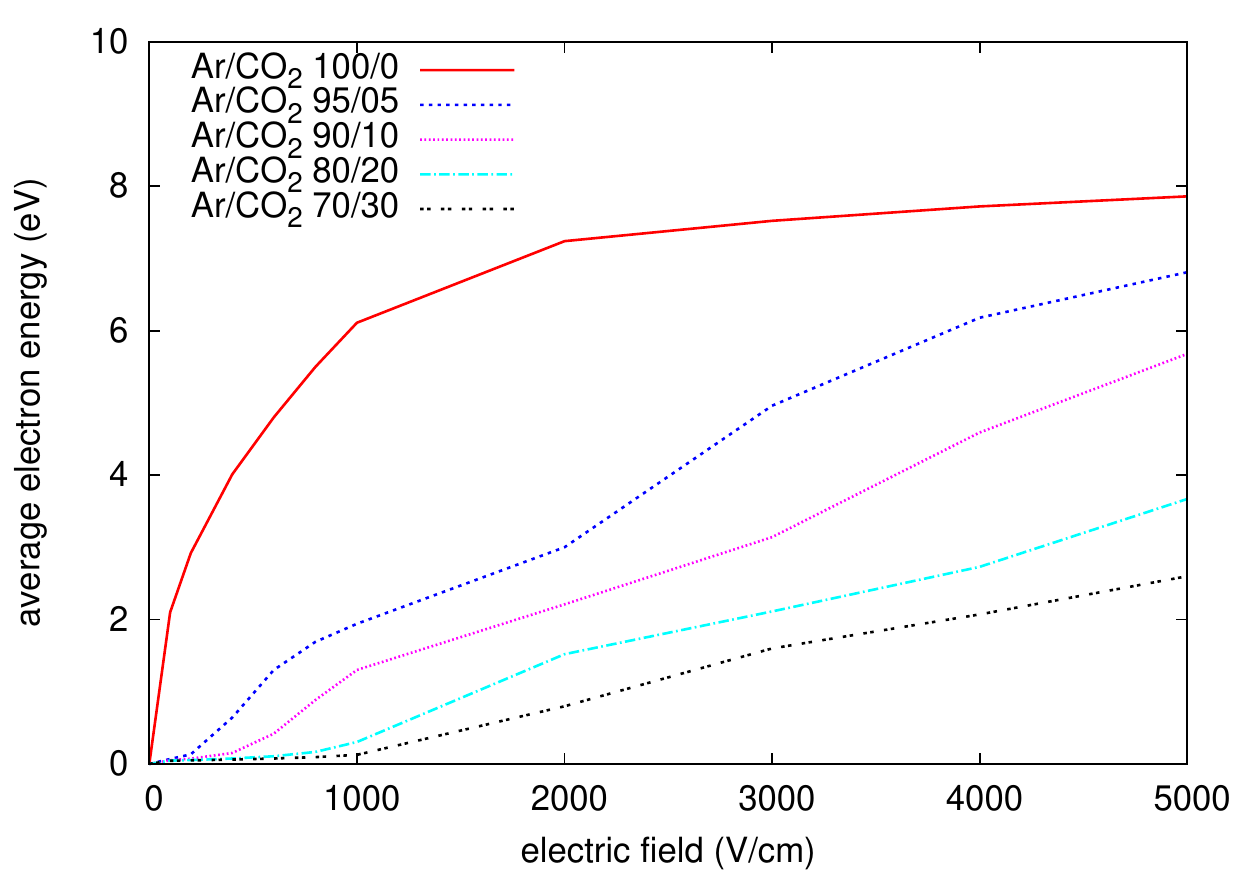}
\caption{Average electron energies for different mass ratios of Ar/CO$_2$.}
\label{fig:electronEnergyAverage}
\end{figure}

As shown in Fig. \ref{fig:electronEnergyAverage} for the gas mixture and drift fields typically used, the average electron energy is lower than 2~eV. 
Increasing the Argon content or moving to Neon based mixtures helps increasing the electron energy and therefore the graphene transparency.
The electron energy increases also with the field, for this reason graphene triple layers were successfully transferred onto the bottom electrode of GEM where the field exceeds 70~kV/cm.
Finally, the electron transparency for a wider range of electron energies will be investigated by systematic measurements with an electron point source in vacuum.

If the strong asymmetry between electron and ion transfer can be reproduced at low energies and in the presence of gas, the graphene can have important applications in reducing the ion back-flow into the conversion region of gas detectors, yielding, for instance, an improvement of distortions induced by space charges in Time Projection Chambers and reducing the ion feedback from the cathode.
Graphene layers on the bottom electrode of the top GEM will effectively block the ion backflow from amplification stages below this device, leaving only negligible ion backflow from the top GEM as first amplification even for high rate measurements.\\

\section*{Acknowledgement}
T.T.N. and R.B.J. thank Applied Scintillation Technologies Ltd for financial support, including the award to T.T.N. of a PhD studentship.
We would like to thank Barbora Bartova (CERN, Switzerland) for her support on SEM imaging.

\end{document}